\documentclass[11pt,journal,draftcls,letterpaper,onecolumn]{IEEEtran}

\usepackage{amsmath, amsthm, amssymb,graphicx,subfigure,algorithm,algorithmic}
\usepackage{times, cite,multirow}
\begin{document}
\title{Wireless Network Coding with Imperfect Overhearing}
\author{Haishi~Ning,~\IEEEmembership{Student Member,~IEEE,} Cong~Ling,~\IEEEmembership{Member,~IEEE,} and~Kin~K.~Leung,~\IEEEmembership{Fellow,~IEEE}%
\thanks{The work reported in this paper has formed part of the Flexible Networks area of the Core 5 Research Programme of the Virtual Center of Excellence in Mobile \& Personal Communications, Mobile VCE, www.mobilevce.com, and has been jointly funded by Mobile VCE's industrial member companies and the UK Government, via the Engineering and Physical Sciences Research Council.}%
\thanks{This work was published in part in the Second IEEE International Workshop on Wireless Network Coding (WiNC 2009), Rome, Italy, June 2009.}%
\thanks{The authors are with Department of Electrical and Electronic Engineering, Imperial College London, London SW7 2AZ, UK (e-mail: haishi.ning06@imperial.ac.uk; c.ling@imperial.ac.uk; kin.leung@imperial.ac.uk).}}
\maketitle
\begin{abstract}
\label{abstract}
Not only is network coding essential to achieve the capacity of a single-session multicast network, it can also help to improve the throughput of wireless networks with multiple unicast sessions when overheard information is available. Most previous research aimed at realizing such improvement by using perfectly overheard information, while in practice, especially for wireless networks, overheard information is often imperfect. To date, it is unclear whether network coding should still be used in such situations with imperfect overhearing. In this paper, a simple but ubiquitous wireless network model with two unicast sessions is used to investigate this problem. From the diversity and multiplexing tradeoff perspective, it is proved that even when overheard information is imperfect, network coding can still help to improve the overall system performance. This result implies that network coding should be used actively regardless of the reception quality of overheard information.
\end{abstract}
\section{Introduction}
\label{introduction}
Network coding was initially proposed in \cite{ahlswede} to achieve the capacity of a single-session multicast network by permitting intermediate nodes to encode received data rather than just to do traditional routing operations. For a single-session multicast network, it was shown in \cite{lnc} that linear codes are sufficient to achieve the multicast capacity. A polynomial time algorithm for network code construction was proposed in \cite{jaggi}. Later, a distributed random linear code construction approach was proposed in \cite{medard}, which was also shown to be asymptotic valid given a sufficiently large field size. For a multiple-session network, it was shown in \cite{riis,dougherty} that linear network coding may be insufficient to achieve the capacity. Moreover, finding a network coding solution for a network with multiple sessions was shown to be a NP-hard problem \cite{koetter,lehman}.

Although optimal network coding solutions for multiple-session networks are generally unknown, simple network coding solutions are able offer tremendous throughput improvement, which was famously demonstrated by \cite{wuyunnan1,cope,pnc}. In those works, the information overheard by a node or previously transmitted by a node, that can be used in the decoding process, is either perfect or ignored. While it is reasonable to assume a node's previously transmitted information to be perfect, it is less so to assume the overheard information to be lossless, especially in wireless networks with fading and noise corruption. In the situations with imperfectly overheard information, previous research often ignored it and used traditional routing solutions instead.

One may naturally ask whether network coding should still be used if the overheard information is imperfect. This is a common and important problem because it is clearly wasteful to ignore the whole overheard information, while only a few symbols in it are incorrect. On the other hand, with imperfect overhearing, a node may not be able to ``naively'' decode its desired information by simply removing the interference which is related to the overheard information.
\subsection{Problem formulation}
\begin{figure}
    \centering
    \includegraphics[width=12cm]{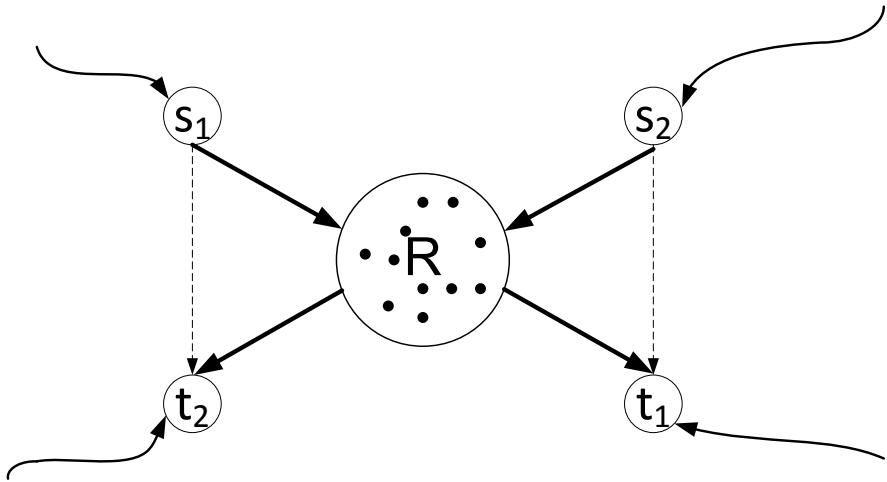}
    \caption{Channel model with two source nodes, two destination nodes, and a cluster of $N$ relay nodes.}
    \label{fig:channelmodel}
\end{figure}
Consider the problem as shown in Fig. \ref{fig:channelmodel}. It is an abstraction of a practical wireless two-session two-hop relay channel which could exist in a wireless mesh network with rich overhearing opportunities and cheap but dense relays. In this channel model, two source nodes $\mathcal{S}=\{s_1,s_2\}$ want to transmit independent information to two destination nodes $\mathcal{T}=\{t_1,t_2\}$ respectively. Due to the distance between the sources and their corresponding destinations, each source has to transmit its information to the intended destination through the help of a cluster of $N$ relay nodes $\mathcal{R}=\{r_1,r_2,...,r_N\}$.

This channel model is ubiquitous in practice because the sources and destinations do not have to be the true communication end-users. It can happen as long as two traditional routing paths intersect at some point and share one or more intermediate relays. Because of the shared use of the resources, there is higher throughput and reliability requirements at the shared relays. This motivates us to develop new transmission strategies to meet the ongoing higher and higher quality-of-service (QoS) requirements.

Many transmission strategies have been proposed for this channel model. Each of them has its own advantages and drawbacks. Some are aimed at achieving maximum diversity gain and some are aimed at achieving maximum multiplexing gain. When comparing different transmission strategies, we want to use the diversity and multiplexing tradeoff (DMT) as a fundamental benchmark, which can characterize both throughput and robustness at the same time \cite{zhenglizhong}. A well known analogy is that a code with longer redundancy may have more powerful error correcting ability. However, its closeness to the Shannon limit is a more fundamental measurement than either its rate or error correcting ability alone.
\subsubsection{Traditional multihop routing}
\begin{figure}
    \centering
    \includegraphics[width=14cm]{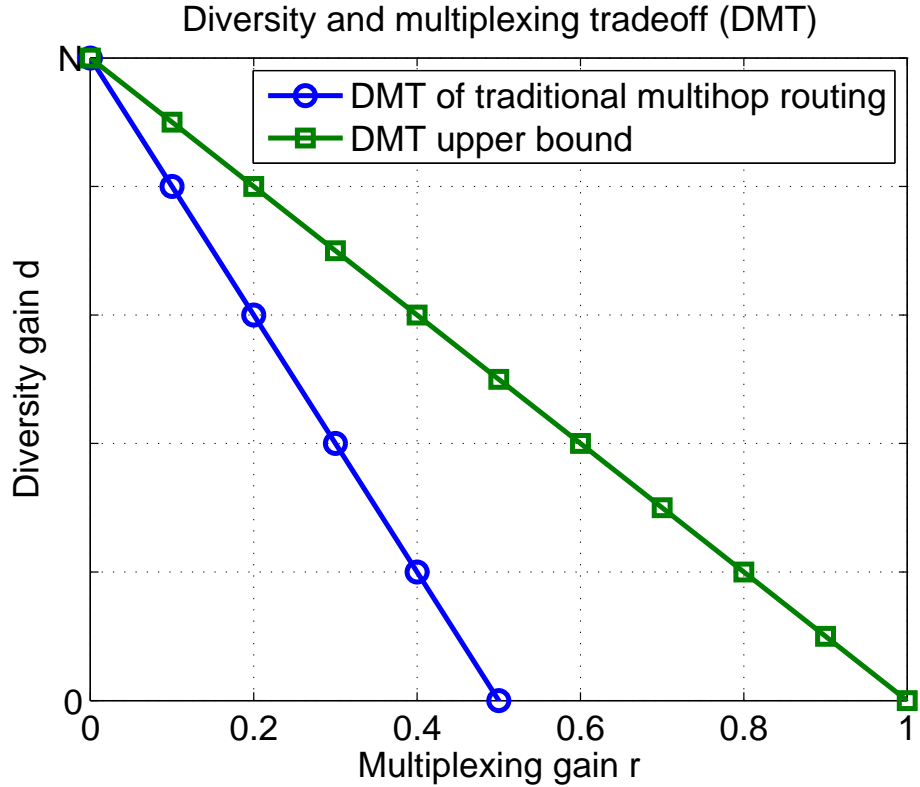}
    \caption{DMT of traditional multihop routing strategy.}
    \label{fig:tmt}
\end{figure}
Traditional multihop routing strategy transmits information over multiple hops along paths from the sources to the destinations. It uses only point-to-point coding, treating all interference as noise and the information is fully decoded at each intermediate relay. Much of current protocol development activity is based on the idea of multihop routing. From the transport capacity point of view, several network information theorists have justified the order optimality of multihop routing in the relatively high attenuation scenario \cite{gupta1,gupta2,xieliangliang}. This order optimality of the transport capacity characterizes the achievable throughput in the error-free case. In practice, the slope of the bit error rate (BER) is also important because we want to set up the communication with some acceptable QoS, and thus the DMT characteristic is also an important measurement.

Using traditional multihop routing strategy, the idea of interference avoidance is often used to achieve an acceptable QoS. Thus, we need four time slots to complete the communication task, i.e., each source uses one time slot to transmit its information to the relays. The relays fully decode each source's information and then forward each of them to its intended destination using one time slot respectively. This multihop routing strategy is indeed a realization of the decode and forward (DF) strategy as shown in \cite{laneman2}. For clarity, we show DMT for this strategy, which is the same as that of DF strategy in Fig. \ref{fig:tmt}.
\subsubsection{Digital network coding}
\begin{figure}
    \centering
    \includegraphics[width=14cm]{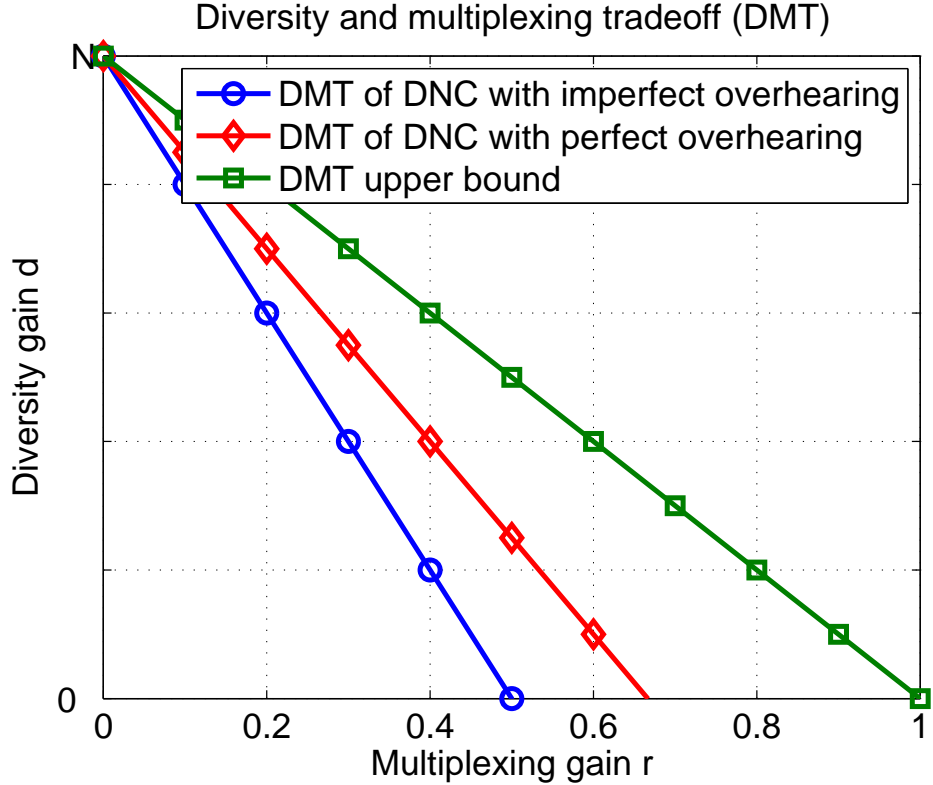}
    \caption{DMT of digital network coding strategy.}
    \label{fig:tnc}
\end{figure}
Consider the transmissions from the sources to the relays as the first phase and the transmissions from the relays to the destinations as the second phase.

The first phase of digital network coding (DNC) strategy is just like that of traditional multihop routing strategy, where each source transmits its information to the relays sequentially. Moreover, due to the wireless broadcast nature, each source's unintended destination can also receive the signal emitted by the undesired source, i.e., $t_1$ can receive signal from $s_2$ and $t_2$ can receive signal from $s_1$.

Firstly, we assume each destination can perfectly decode this overheard information and stores it in its memory stack. Then, the second phase of DNC strategy involves the exclusive-or (XOR) between the two sources' information at the relays. Again, due to the wireless broadcast nature, when the relays broadcast the network coded information, both destinations can receive the signal. After decoding the XOR of the two sources' information, each destination XORs it again with its previously stored overheard information, in order to extract its desired information. Thus, it only needs three time slots to complete the communication task using DNC strategy with perfect overhearing. The DMT characteristic of DNC in this case is similar to that of DF strategy with improved efficiency (multiplexing gain).

Secondly, we consider the situation with imperfect overhearing, i.e., when one or both of the destinations cannot perfectly decode the overheard information. In this case, the destinations simply discard the imperfectly overheard information. Thus, since there is not enough overheard information to help in the decoding process, the relays cannot broadcast network coded packets to the destinations. Instead, with imperfect overhearing, DNC strategy falls back to traditional multihop routing strategy and still uses four time slots to complete the communication task. The complete DMT characteristic of DNC strategy is shown in Fig. \ref{fig:tnc}.
\subsubsection{Physical-layer network coding}
\begin{figure}
    \centering
    \includegraphics[width=14cm]{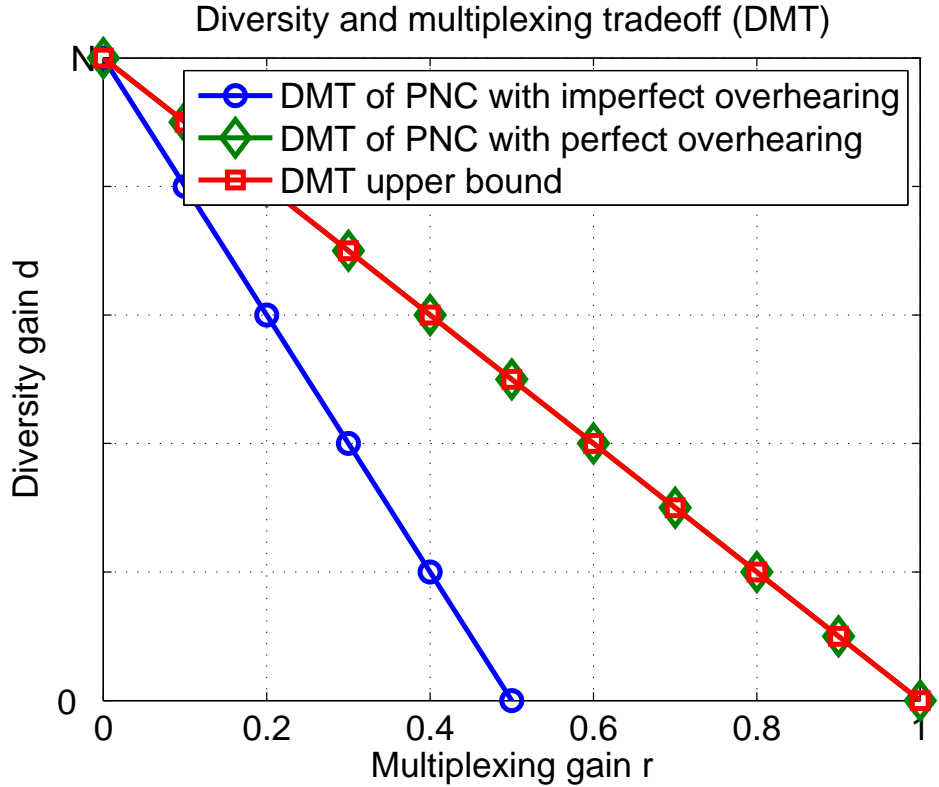}
    \caption{DMT of physical-layer network coding strategy.}
    \label{fig:pnc}
\end{figure}
In \cite{pnc,anc}, it was shown that the network coding operations at the relays in DNC can also be done in the air using electromagnetic waves. The second phase of physical-layer network coding (PNC) strategy is just like that of DNC, i.e., in the case with perfect overhearing, the relays broadcast the network coded information to both destinations in one time slot due to the wireless broadcast nature; in the case with imperfect overhearing, PNC also falls back to traditional multihop routing strategy and uses two time slots in the second phase.

If the overheard information can be perfectly decoded by both destinations, then two sources can transmit their information simultaneously to the relays in the first phase of PNC. Instead of decoding each source's information separately, the relays decode the XOR of two sources' information directly using the superpositioned signal. Thus, PNC saves one time slot compared to DNC in the first phase and only needs two time slots to complete the entire communication task with perfect overhearing. However, if the overheard information is imperfect, then the relays need to transmit each source's information separately to the corresponding destination. Thus, PNC still needs two time slots for each source to transmit its information to the relays in the first phase with imperfect overhearing, and thus four time slots to complete all. In summary, with perfect overhearing, PNC has the same DMT characteristic as that of amplify and forward strategy in \cite{laneman1}, and otherwise, PNC essentially falls back to multihop routing. The DMT characteristic of PNC strategy is illustrated in Fig. \ref{fig:pnc}.
\subsection{Motivation and objective}
From the arguments in last few subsections, it is clear that state-of-the-art implementations of both DNC and PNC are subject to the reception of perfectly overheard information at the destinations, i.e., $t_1$ and $t_2$ have to be able to fully decode the information sent by $s_2$ and $s_1$ respectively. If the overhearing is imperfect, then they have to fall back to traditional multihop routing strategy. They usually ignore the imperfectly overheard information, although in symbol level, the corrupted overheard packets can also be helpful in the decoding process at the destination nodes.

A question one may naturally ask is: when the overheard information is imperfect, what can we do to increase the overall system performance? Or equivalently, can we use the imperfectly overheard information to improve the network throughput or robustness? The answer is important and instructive to researchers even in other fields. If the answer is ``No", then we have to continue our protocol development activities by using currently prevalent multihop routing strategy or an adaptive strategy which can automatically switch its operation mode between the perfect and imperfect overhearing situations; if the answer is ``Yes", then we can potentially increase the overall system performance where state-of-the-art strategies are incapable to do.

Whether we can use the imperfectly overheard information to improve the overall system performance is the subject we want to study in this paper. We show that even when the overheard information is imperfect, the overall system performance can still be improved if network coding is used actively.
\section{Active physical-layer network coding}
Consider the problem as shown in Fig. \ref{fig:channelmodel}. In order to show network coding should be used actively to improve the overall system performance even with imperfect overhearing, we use an active PNC strategy so that physical-layer network coding is applied actively regardless of the reception quality of the overheard information. Thus, we evaluate the performance of the following transmission strategy:
\begin{enumerate}
\item There are two time slots in each transmission frame.
\item In the first time slot, two source nodes $s_1$ and $s_2$ broadcast their independent information $x_{s_1}$ and $x_{s_2}$ simultaneously to the $N$ shared relay nodes and their unintended destination nodes respectively. The destination nodes try to decode the overheard information and store the decoded information in its own memory stack.
\item In the second time slot, the $N$ shared relay nodes normalize their received signal in the first time slot and broadcast the normalized signal simultaneously to the two destination nodes. The destination nodes use the received signal from the relays and together with the stored information in the first time slot to extract their desired information.
\end{enumerate}

The main difference between this active PNC strategy and traditional PNC strategy is its ``activeness'', which means the relay nodes actively and intentionally mix the two sources' signals no matter whether the overheard information is perfect or not. Conventional DNC and PNC fall back to multihop routing when the overheard information is imperfect, even when there are only a few incorrectly overheard symbols. We will show that even when the overhearing is imperfect, it can still help to improve the overall system performance, which indicates that network coding should be applied actively regardless of the reception quality of the overheard information.
\subsection{Practical considerations}
\subsubsection{Channel side information}
We assume channel side information is only available at the receivers (CSIR), which is practicable by inserting a negligibly short training sequence into the message sequences. Moreover, we let the relay nodes broadcast their estimated CSIR and normalization factors by embedding them into the training sequence with negligible overhead compared to the original message length.
\subsubsection{Synchronization}
We do not consider the synchronization issue and assume all the relay nodes are fully synchronized. The synchronization issue arising from the simultaneous relaying operations can be overcome by a distributed relay selection algorithm which chooses only the best relay node to forward its received signal from the source nodes in the first time slot. Moreover, from \eqref{eqn:outageorder1} and \eqref{eqn:outageorder2} stated later, the performance of simultaneous relaying is dominated by the best two-hop link between the sources and the corresponding destinations. Thus, a distributed relay selection algorithm does not entail a cost on the achievable DMT. However, we only consider simultaneous relaying in this paper for mathematical simplicity. The same asymptotic DMT performance can be achieved with a suitable  distributed relay selection algorithm such as that in \cite{dingzhiguo2}.
\subsection{Decoding at the destinations}
\begin{figure}
    \centering
    \includegraphics[width=17cm]{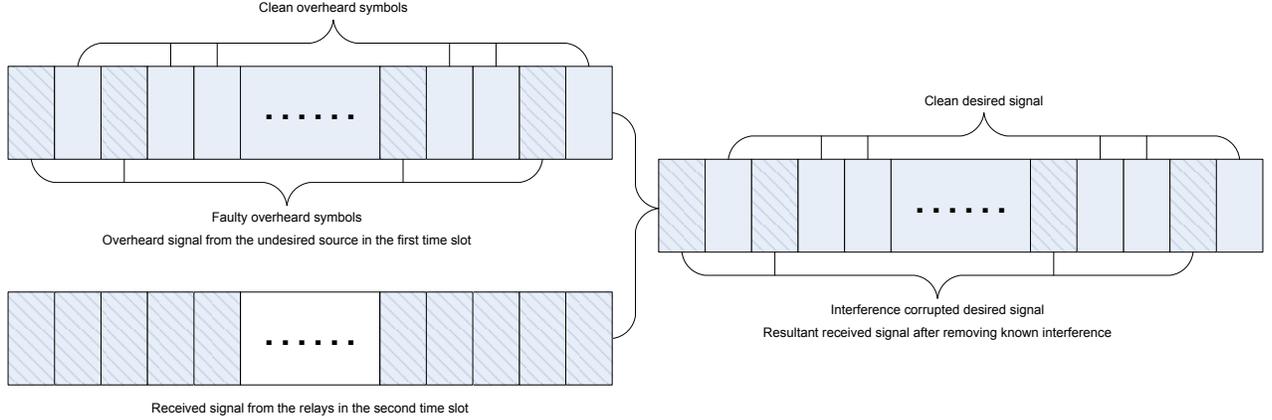}
    \caption{Decoding at the destination nodes.}
    \label{fig:44}
\end{figure}
We assume all the physical links are quasi-static flat Rayleigh fading. For the source-to-relay and relay-to-destination links that suffer deep fading, it is feasible to use some physical layer error correction codes or even higher layer protocols to ensure they are error-free because they are the "designed" transmissions. However, for the overheard links, it is infeasible in practice to spend extra resource to ensure their reliability. Thus, when the destinations try to decode their overheard information in the first time slot, the decoded packets may be imperfect with symbol errors. We divide the decoded overheard information at the destinations in the first time slot into two parts: one part with high probability to be correct (clean overheard symbols); the other part with high probability to be incorrect (faulty overheard symbols). Possible methods to mark the decoded symbols as clean or faulty can use the output of many well known soft decoders or the confidence values calculated from the physical layer signal as shown in \cite{hanzobook,jamieson}. The comparison between different marking methods and their associated error propagation effects are out of the scope of this paper. The purpose for such division is that we want to use the clean overheard symbols to remove part of the interference in the received signal in the second time slot.

When each destination receives the signal from the relays in the second time slot, it uses the clean overheard symbols together with the channel side information and the normalization factors to reconstruct part of the interference. Then, it subtracts such reconstructed interference from its received signal in the corresponding positions. This results in dividing the desired signal at the destinations into two parts: one part without interference and the other part with unknown interference. For the second part with unknown interference signal, we will use traditional decoding method for multiple-access channels (MAC) to extract the desired information. Fig. \ref{fig:44} illustrates the decoding process at the destinations.
\subsection{Preliminaries}
Throughout this paper, we use $\mathcal{S}=\{s_1,s_2\}$ to denote the two source nodes, $\mathcal{T}=\{t_1,t_2\}$ to denote the two destination nodes and $\mathcal{R}=\{r_1,r_2,...,r_N\}$ to denote the $N$ relay nodes. We use $x_{s_1}$ and $x_{s_2}$ to denote the signal transmitted from the two source nodes and $x_{r_n}$ to denote the signal transmitted from the $n$th relay node, for $1\leqslant{n}\leqslant{N}$. Similarly, $y_{r_n}$ is used to represent the received signal at the $n$th relay node and $y_{t_1}$ and $y_{t_2}$ are used to denote the received signal at the two destination nodes respectively.

Every node is constrained by average energy $E$. All source nodes transmit independent information at the same rate $R$. $h_{s_m,r_n}$, $h_{s_m,t_k}$ and $h_{r_n,t_k}$ are used to denote the channel gain between the $m$th source node and the $n$th relay node, the channel gain between the $m$th source node and the $k$th destination node and the channel gain between the $n$th relay node and the $k$th destination node, where $1\leqslant{m}\neq{k}\leqslant{2}$ and $1\leqslant{n}\leqslant{N}$. We assume the existing physical links are all quasi-static flat Rayleigh-fading, which means the channel gains are constant during each frame but change independently between different frames.

We characterize the channels between the sources and their unintended destination nodes using the amount of information the destinations overheard from their undesired sources in the first time slot. In order to evaluate the performance using active physical-layer network coding when the overheard information is imperfect, we assume that in the first time slot, each destination node can only decode part of its undesired source's information correctly, i.e., $t_1$ can decode $R_{t_1}$ amount of information from $s_2$ correctly, and $t_2$ can decode $R_{t_2}$ amount of information from $s_1$ correctly, where $0\leqslant{R_{t_1},R_{t_2}}\leqslant{R}$. From \cite{coverbook}, we know this amount of information can be translated into each codeword's bin index, i.e., $2^{R_{t_1}}$ out of $2^R$ bits for $x_{s_1}$ and $2^{R_{t_2}}$ out of $2^R$ bits for $x_{s_2}$.

The mathematical tools we use are mainly from \cite{zhenglizhong,laneman1,laneman2,azarian,shengyang,telatar}. For clarity, we state some background knowledge as follows:
\begin{enumerate}
\item The transmit SNR of a physical link is defined as $\rho=\frac{E}{\sigma^2}$ where $E$ is the average signal energy at the transmitter and $\sigma^2$ is the AWGN variance at the receiver. We say $b$ is the exponential order of $f(\rho)$ \cite{zhenglizhong} if
\begin{eqnarray}
\lim_{\rho\rightarrow\infty}\frac{\log(f(\rho))}{\log(\rho)}=b
\end{eqnarray} and denote $f(\rho)$ as $f(\rho)\dot{=}\rho^b$. $\dot{\leqslant}$ and $\dot{\geqslant}$ are similarly defined.
\item Consider a coding scheme as a family of codes $\{C(\rho)\}$ with data rate $R(\rho)$ bits per channel use (BPCU) and average maximum-likelihood (ML) error probability $P_E(\rho)$. The multiplexing gain $r$ and the diversity gain $d$ are defined as \cite{zhenglizhong}
\begin{eqnarray}
r=\lim_{\rho\rightarrow\infty}\frac{R(\rho)}{\log(\rho)},\quad d=-\lim_{\rho\rightarrow\infty}\frac{\log(P_E(\rho))}{\log(\rho)}.
\end{eqnarray}
\item Let $h$ be complex standard normal distributed and $v$ denotes the exponential order of $\frac{1}{|h|^2}$. The probability density function (pdf) of $v$ can be written as \cite{azarian}
\begin{eqnarray}
p_v\dot{=}
  \left\{
    \begin{array}{l l}
      \rho^{-\infty}=0, & \textrm{for} \:v<0\\
      \rho^{-v}, & \textrm{for} \:v\geqslant{0}.\\
    \end{array}
    \right.
\end{eqnarray} Thus, for $N$ independent identically distributed (i.i.d) variables $\{v_j\}_{j=1}^N$, the probability that $(v_1,...,v_N)$ belongs to a set $O$ is
\begin{eqnarray}
P_O\dot{=}\rho^{-d_O}, \quad \textrm{ for } \: d_O=\inf_{(v_1,...,v_N)\in{O^+}}\sum_{j=1}^Nv_j
\label{eqn:do}
\end{eqnarray} given that $\mathbb{R}^{N+}$ denotes the set of nonnegative $N$-tuples and $O^+=O\bigcap{}\mathbb{R}^{N+}$ is not empty. Thus, the exponential order of $P_O$ depends only on $O^+$ and is dominated by the realization with the largest exponential order.
\item Let $d_1=1-Ar_1$ and $d_2=1-Br_2$ be two linear functions that denote the DMTs of two independent messages, where $A$ and $B$ are two constants. The overall DMT is obtained by adding the multiplexing gains up subject to equal diversity gains, and can be written as
    \begin{eqnarray}
    d=1-\frac{AB}{A+B}r.
    \label{eqn:overalldmt}
    \end{eqnarray}
\end{enumerate}
\subsection{Performance evaluation}
\subsubsection{Signaling and DMT analysis for the part of the signal with known interference}
Since there is no difference in processing different symbols, we assume the first part of the received signal at the $n$th relay node $r_n$ is a combination of two super-symbols. In the first time slot, the received signal vector at the $N$ relay nodes is
\begin{eqnarray}
\textbf{y}_{\mathcal{R},N\times{1}} &=& \textbf{H}_{\mathcal{S,R},N\times2}\cdot{}\textbf{x}_{\mathcal{S},2\times1}+\textbf{n}_{\mathcal{R},N\times{1}} \notag \\
&=&
\begin{pmatrix}
h_{s_1,r_1} & h_{s_2,r_1} \\
h_{s_1,r_2} & h_{s_2,r_2} \\
\multicolumn{2}{c}{...} \\
h_{s_1,r_N} & h_{s_2,r_N}
\end{pmatrix}_{N\times2}
\cdot{}
\begin{pmatrix}
x_{s_1} \\
x_{s_2}
\end{pmatrix}_{2\times1}+
\begin{pmatrix}
n_{r_1} \\
n_{r_2} \\
... \\
n_{r_N}
\end{pmatrix}_{N\times1}.
\label{eqn:relayrx}
\end{eqnarray}

In the second time slot, the signal transmitted from the $n$th relay node is
\begin{eqnarray}
x_{r_n}=\beta_{r_n}\cdot{}y_{r_n} \text{, for } n=1,2,...,N,
\label{eqn:relaytx}
\end{eqnarray}
where $\beta_{r_n}$ is the normalization factor at the relay node $r_n$ which is used to ensure the relay node to satisfy its average energy constraint $E$. Each relay node chooses its energy normalization factor $\beta_{r_n}$ based on its own received signal energy. We use an equal power allocation scheme with the average energy constraint $E$ for each relay node in the second time slot of one transmission frame. A more advanced power allocation scheme will enhance the performance in terms of the throughput and outage probability only in the low SNR regime. However, such improvement becomes trivial in the high SNR regime and a simple equal power allocation scheme is sufficient to achieve the same DMT as that of the optimal power allocation scheme.

The received signal at the destination nodes in the second time slot can be written as
\begin{eqnarray}
y_{t_1}=\textbf{H}_{\mathcal{R},t_1,1\times{N}}\cdot{}\textbf{x}_{\mathcal{R},N\times1}+n_{t_1}
\label{eqn:destinationrx}
\end{eqnarray}
where $\textbf{H}_{\mathcal{R},t_1,1\times{N}}=[h_{r_1,t_1},h_{r_2,t_1},...,h_{r_N,t_1}]$, $\textbf{x}_{\mathcal{R},N\times1}=[x_{r_1},x_{r_2},...x_{r_N}]^\dag$ and $[\cdot]^\dag$ denotes the matrix conjugated transposition. From \eqref{eqn:relayrx} and \eqref{eqn:relaytx}, we know $\textbf{x}_{\mathcal{R},N\times1}$ can also be written as
\begin{eqnarray}
\textbf{x}_{\mathcal{R},N\times1}&=&\textbf{$\beta$}_{N\times{N}}\cdot{}\textbf{y}_{\mathcal{R},N\times{1}} \notag \\
&=&
\begin{pmatrix}
\beta_{r_1} & 0 & 0 & ... & 0 \\
0 & \beta_{r_2} & 0 & ... & 0 \\
0 & 0 & \beta_{r_3} & ... & 0 \\
\multicolumn{5}{c}{...} \\
0 & 0 & 0 & ... & \beta_{r_N}
\end{pmatrix}_{N\times{N}}\cdot{}
\begin{pmatrix}
y_{r_1} \\
y_{r_2} \\
y_{r_3} \\
...\\
y_{r_N}
\end{pmatrix}_{N\times1}.
\label{eqn:relaytx2}
\end{eqnarray}
Substitute \eqref{eqn:relayrx} and \eqref{eqn:relaytx2} into \eqref{eqn:destinationrx}, we can get
\begin{eqnarray}
y_{t_1} &=& \textbf{H}_{\mathcal{R},t_1,1\times{N}}\cdot{}\textbf{x}_{\mathcal{R},N\times1}+n_{t_1}=\textbf{H}_{\mathcal{R},t_1,1\times{N}}\cdot{}\textbf{$\beta$}_{N\times{N}}\cdot{}\textbf{y}_{\mathcal{R},N\times{1}}+n_{t_1}\notag\\
&=&\textbf{H}_{\mathcal{R},t_1,1\times{N}}\cdot{}\textbf{$\beta$}_{N\times{N}}\cdot{}(\textbf{H}_{\mathcal{S,R},N\times2}\cdot{}\textbf{x}_{\mathcal{S},2\times1}+\textbf{n}_{\mathcal{R},N\times{1}})+n_{t_1}\notag\\
&=& [h_{r_1,t_1},h_{r_2,t_1},......,h_{r_N,t_1}]_{1\times{N}}\notag \\
&\quad&\cdot{}
\begin{pmatrix}
\beta_{r_1}& 0 & 0 & ... & 0 \\
0 & \beta_{r_2} & 0 & ... & 0 \\
0 & 0 & \beta_{r_3} & ... & 0 \\
\multicolumn{5}{c}{...} \\
0 & 0 & 0 & ... & \beta_{r_N}
\end{pmatrix}_{N\times{N}}\cdot
\begin{pmatrix}
h_{s_1,r_1} & h_{s_2,r_1} \\
h_{s_1,r_2} & h_{s_2,r_2} \\
\multicolumn{2}{c}{...} \\
h_{s_1,r_N} & h_{s_2,r_N}
\end{pmatrix}_{N\times2} \cdot
\begin{pmatrix}
x_{s_1} \\
x_{s_2}
\end{pmatrix}_{2\times1}
+\tilde{n}_{t_1}.
\label{eqn:destinationrx1}
\end{eqnarray}
Thus,
\begin{eqnarray}
y_{t_1}&=&\sum_{n=1}^Nh_{s_1,r_n}\beta_{r_n}h_{r_n,t_1}x_{s_1}+\sum_{n=1}^Nh_{s_2,r_n}\beta_{r_n}h_{r_n,t_1}x_{s_2} \notag \\
&\quad&+(\sum_{n=1}^Nh_{r_n,t_1}\beta_{r_n}n_{r_n}+n_{t_1}).
\label{eqn:destinationrx2}
\end{eqnarray}

With CSIR and normalization factors received from the relay nodes and estimated by the destination nodes themselves, each destination node can remove its known interference from its received signal in the second time slot. For destination node $t_1$, $\sum_{n=1}^Nh_{s_2,r_n}\beta_{r_n}h_{r_n,t_1}x_{s_2}$ is the known signal and thus can be removed from \eqref{eqn:destinationrx2}. Thus, we can write
\begin{eqnarray}
y_{t_1}=\sum_{n=1}^Nh_{s_1,r_n}\beta_{r_n}h_{r_n,t_1}x_{s_1}+(\sum_{n=1}^Nh_{r_n,t_1}\beta_{r_n}n_{r_n}+n_{t_1}).
\label{eqn:destinationrx3}
\end{eqnarray}

The accumulated noise at the first destination node $t_1$ from both relay and destination nodes can be written as
\begin{eqnarray}
\tilde{n}_{t_1}=\sum_{n=1}^Nh_{r_n,t_1}\beta_{r_n}n_{r_n}+n_{t_1}
\label{eqn:accumulatednoise}
\end{eqnarray}
where the normalization factor $\beta_{r_n}$ is chosen to satisfy energy constraint
\begin{eqnarray}
|\beta_{r_n}|^2&\leqslant{}&\frac{E}{E|h_{s_1,r_n}|^2+E|h_{s_2,r_n}|^2+\sigma^2}\notag \\
&=&\frac{\rho}{\rho|h_{s_1,r_n}|^2+\rho|h_{s_2,r_n}|^2+1}.
\label{eqn:betaconstraint}
\end{eqnarray}
where $\sigma^2$ is the noise variance.

Let $w_n$ denote the exponential order of $|\beta_{r_n}|^2$ and $v_{i,n}$ and $u_{n,j}$ denote the exponential orders of $\frac{1}{|h_{s_i,r_n}|^2}$ and $\frac{1}{|h_{r_n,t_j}|^2}$ respectively, for $i,j=1,2$ and $1\leqslant{n}\leqslant{N}$. Thus, from \eqref{eqn:betaconstraint}, we can easily see that
\begin{eqnarray}
w_{n}\leqslant{}\min{(v_{1,n},v_{2,n},1)}.
\label{eqn:betaorderconstraint}
\end{eqnarray}
For \eqref{eqn:betaorderconstraint} to be met, we choose $w_{n}$ as
\begin{eqnarray}
w_{n}=(v_{1,n},v_{2,n})^-
\label{eqn:betachoice}
\end{eqnarray}
where we use $(x)^-$ to mean $\min\{x,0\}$ and $(x)^+$ to mean $\max\{x,0\}$. This choice for $w_{n}$ will ensure $\beta_{r_n}$ to satisfy the energy constraint \eqref{eqn:betaconstraint}. This, under the consideration of outage events belonging to set $\emph{O}^+$ as stated in \eqref{eqn:do} will make $w_{n}$, i.e., the exponential order of $\beta_{r_n}$ vanish in all the DMT analytical expressions.

Let $w_{\tilde{n}_{t_1}}$ and $w_{n_{t_1}}$ denote the exponential orders of the variances of $\tilde{n}_{t_1}$ and $n_{t_1}$. From \eqref{eqn:accumulatednoise} and \eqref{eqn:betachoice}, we know
\begin{eqnarray}
w_{\tilde{n}_{t_1}}=\max_{n=1,2,...,N}\{(-u_{n,1})^+\}+w_{n_{t_1}}=w_{n_{t_1}}.
\label{eqn:noisedmt}
\end{eqnarray}

Thus, the DMT of the active PNC strategy depends only on the channel matrix and not on the variance of the accumulated noise. So, for analytical simplicity, we assume the accumulated noise equals to the noise at each destination node which does not affect the DMT analysis.
Thus, we can rewrite \eqref{eqn:destinationrx3} as
\begin{eqnarray}
y_{t_1}=\sum_{n=1}^Nh_{s_1,r_n}\beta_{r_n}h_{r_n,t_1}x_{s_1}+n_{t_1}.
\label{eqn:yd1final}
\end{eqnarray}
and $y_{t_2}$ can be similarly written as
\begin{eqnarray}
y_{t_2}=\sum_{n=1}^Nh_{s_2,r_n}\beta_{r_n}h_{r_n,t_2}x_{s_2}+n_{t_2}.
\label{eqn:yd2final}
\end{eqnarray}

Observing \eqref{eqn:yd1final} and \eqref{eqn:yd2final}, we immediately notice that they are very similar to multiple-input single-output (MISO) channels and thus should have similar DMT characteristics. We obtain a lower bound of the DMT of the active PNC strategy by firstly approximating the exponential order of the error probability of ML decoder by that of the outage probability. From the definition of the outage probability, we know that
\begin{eqnarray}
P_{O_1}&=&P[I(x_{s_1};y_{t_1}|x_{s_2})<R_{t_1}] \notag\\
&=&P[\log(1+\rho\sum_{n=1}^N|h_{s_1,r_n}|^2|\beta_{r_n}|^2|h_{r_n,t_1}|^2)<r_{t_1}\log\rho]
\label{eqn:outage1}
\end{eqnarray}
and
\begin{eqnarray}
P_{O_2}&=&P[I(x_{s_2};y_{t_2}|x_{s_1})<R_{t_2}] \notag\\
&=&P[\log(1+\rho\sum_{n=1}^N|h_{s_2,r_n}|^2|\beta_{r_n}|^2|h_{r_n,t_2}|^2)<r_{t_2}\log\rho].
\label{eqn:outage2}
\end{eqnarray}
In the high SNR regime, the exponential order of $\beta_{r_n}$ vanishes and thus we have
\begin{eqnarray}
&\quad&\lim_{\rho\rightarrow\infty}\frac{I(x_{s_1};y_{t_1}|x_{s_2})}{\log\rho}\notag \\
&=&\lim_{\rho\rightarrow\infty}\frac{\log(1+\rho\sum_{n=1}^N|h_{s_1,r_n}|^2|\beta_{r_n}|^2|h_{r_n,t_1}|^2)}{\log\rho}\notag\\
&=&\max_{n=1,2,...,N}\{1-v_{1,n}-u_{n,1}\}^+
\label{eqn:outageorder1}
\end{eqnarray}
and
\begin{eqnarray}
&\quad&\lim_{\rho\rightarrow\infty}\frac{I(x_{s_2};y_{t_2}|x_{s_1})}{\log\rho}\notag \\
&=&\lim_{\rho\rightarrow\infty}\frac{\log(1+\rho\sum_{n=1}^N|h_{s_2,r_n}|^2|\beta_{r_n}|^2|h_{r_n,t_2}|^2)}{\log\rho}\notag\\
&=&\max_{n=1,2,...,N}\{1-v_{2,n}-u_{n,2}\}^+.
\label{eqn:outageorder2}
\end{eqnarray}
Thus, from \eqref{eqn:outage1}, \eqref{eqn:outage2}, \eqref{eqn:outageorder1} and \eqref{eqn:outageorder2}, the outage events sets $O^+$ should be defined as
\begin{eqnarray}
O^+_1=\{(\textbf{v},\textbf{u})\in\mathbb{R}^{2N+}|\max_{n=1,2,...,N}\{1-v_{1,n}-u_{n,1}\}^+<r_{t_1}\}
\label{eqn:outageevents1}
\end{eqnarray}
and
\begin{eqnarray}
O^+_2=\{(\textbf{v},\textbf{u})\in\mathbb{R}^{2N+}|\max_{n=1,2,...,N}\{1-v_{2,n}-u_{n,2}\}^+<r_{t_2}\}.
\label{eqn:outageevents2}
\end{eqnarray}
From \eqref{eqn:outageevents1} and \eqref{eqn:outageevents2}, we can easily see that, in order for the outage events to happen, the following constraints should be satisfied:
\begin{enumerate}
\item $v_{1,n}+u_{n,1}>1-r_{t_1}, \forall n=1,2,...,N.$
\item $v_{2,n}+u_{n,2}>1-r_{t_2}, \forall n=1,2,...,N.$
\end{enumerate}
From \eqref{eqn:do}, we know the outage probability should be dominated by the probability of the outage event with the largest exponential order, i.e., the outage event with the smallest $d_O(r)$. Thus, we can write
\begin{eqnarray}
P_{O_1}\dot{=}\rho^{-d_{O_1}(r_{t_1})},
\label{eqn:do1}
\end{eqnarray}
for $d_{O_1}(r_{t_1})=\inf_{(\textbf{v},\textbf{u})\in{O^+}}[\sum_{n=1}^N(v_{1,n}+u_{n,1})]$. And
\begin{eqnarray}
P_{O_2}\dot{=}\rho^{-d_{O_2}(r_{t_2})},\notag
\label{eqn:do2}
\end{eqnarray}
for $d_{O_2}(r_{t_2})=\inf_{(\textbf{v},\textbf{u})\in{O^+}}[\sum_{n=1}^N(v_{2,n}+u_{n,2})].$

Thus, we can lower-bound $d_{O_1}(r_{t_1})$ and $d_{O_2}(r_{t_2})$ as
\begin{eqnarray}
d_{O_1}(r_{t_1})\geqslant\inf_{(\textbf{v},\textbf{u})\in{O^+}}[\sum_{n=1}^N(v_{1,n}+u_{n,1})]>N(1-r_{t_1})
\end{eqnarray}
and
\begin{eqnarray}
d_{O_2}(r_{t_2})\geqslant\inf_{(\textbf{v},\textbf{u})\in{O^+}}[\sum_{n=1}^N(v_{2,n}+u_{n,2})]>N(1-r_{t_2}).
\end{eqnarray}
As $d_{O_1}(r_{t_1})$ and $d_{O_2}(r_{t_2})$ also serve as lower bounds for $d_1(r_{t_1})$ and $d_2(r_{t_2})$, we can further write
\begin{eqnarray}
d_1(r_{t_1})>N(1-r_{t_1})
\label{eq:lower1}
\end{eqnarray}
and
\begin{eqnarray}
d_2(r_{t_2})>N(1-r_{t_2}).
\label{eq:lower2}
\end{eqnarray}

Now, we show the DMT of $N(1-r_{t_1})$ and $N(1-r_{t_2})$ are actually also upper bounds for the parts of the received signal with known interference at the two destination nodes with $R_{t_1}$ and $R_{t_2}$ amount of overheard information. Assume in the first time slot, the two source nodes can transmit their independent $R_{t_1}$ amount of $x_{s_1}$ and $R_{t_2}$ amount of $x_{s_2}$ reliably to the $N$ relay nodes. In practice, this cannot be done due to the wireless fading environment and noise corruption. However, this assumption is sufficient to give a DMT upper bound for this part of the received signal. In the second time slot, let the $N$ relays fully cooperate through a genie. Because the two destination nodes cannot cooperate, the best achievable performance is obtained by viewing the transmissions from the relay nodes to the destination nodes in the second time slot as two MISO channels.

Without any cooperation between the destination nodes, the DMT of the two MISO channels are $N(1-r_{t_1})$ and $N(1-r_{t_2})$ respectively. Considering two independent MISO channels together does not increase the diversity gain. This is because the two MISO channels in the second time slot are statistically independent without any cooperation and thus provide no further diversity gain by considering them jointly. Thus, we can get upper bounds as
\begin{eqnarray}
d_1(r_{t_1})<N(1-r_{t_1})
\label{eq:upper1}
\end{eqnarray}
and
\begin{eqnarray}
d_2(r_{t_2})<N(1-r_{t_2}).
\label{eq:upper2}
\end{eqnarray}
Combining \eqref{eq:lower1}, \eqref{eq:lower2}, \eqref{eq:upper1} and \eqref{eq:upper2}, we know the DMT of the active PNC strategy for the interference known part of the signal at the two destinations are
\begin{eqnarray}
d_1(r_{t_1})=N(1-r_{t_1})
\label{eq:dmtfirst}
\end{eqnarray}
and
\begin{eqnarray}
d_2(r_{t_2})=N(1-r_{t_2}).
\label{eq:dmtfirst2}
\end{eqnarray}

In order to get the relationship between the diversity gain $d$ and the multiplexing gain $r$, we need to map points in \eqref{eq:dmtfirst} from a coordinate system with $r_{t_1}$ as $x$-axis to a coordinate system with $r_{}$ as $x$-axis. Because $r_{t_1}=\frac{R_{t_1}}{R}r$, points $(0,N)$ and $(1,0)$ are mapped to points $(0,N)$ and $(\frac{R_{t_1}}{R},0)$. Thus, in the new coordinate system, the DMT \eqref{eq:dmtfirst} changes to
\begin{eqnarray}
d_1(r_{t_1})=N(1-\frac{R}{R_{t_1}}r),
\label{eq:dmt24}
\end{eqnarray}
and similarly we have
\begin{eqnarray}
d_2(r_{t_2})=N(1-\frac{R}{R_{t_2}}r).
\label{eq:dmt25}
\end{eqnarray}
Finally, from \eqref{eqn:overalldmt} and taking the consumption of two time slots in one cooperation frame into consideration, we can get the final overall DMT for the interference known part of the received signal at both destinations as
\begin{eqnarray}
d_I(r_{t_1},r_{t_2})=N[1-(\frac{2R}{R_{t_1}+R_{t_2}})r],
\label{eqn:dmt1}
\end{eqnarray}
where $d_I(r_{t_1},r_{t_2})$ is used to denote the diversity gain of the interference known part of the received signal at both destinations.
\subsubsection{Signaling and DMT analysis for the part of the signal with unknown interference}
For this part of the received signal, as we have no information correctly overheard, we cannot remove the interference term from \eqref{eqn:destinationrx2}. Thus, we have no choice but to use traditional decoding method for MAC.

We first note that from the received signal at both destinations, we can at least extract $I(x_{s_1};y_{t_1})+I(x_{s_2};y_{t_2})$ amount of desired information by treating interference as noise. Moreover, in the high SNR regime, for a 2-user interference channel, the total achievable multiplexing gain is $1$. Because the multiplexing gain for MAC $\{x_{s_1},x_{s_2}\}\rightarrow{}y_{t_1}$ is also $1$, thus we have $\lim_{\rho\rightarrow\infty}\frac{I(x_{s_1},x_{s_2};y_{t_1})}{\log\rho}=\lim_{\rho\rightarrow\infty}\frac{I(x_{s_1};y_{t_1})+I(x_{s_2};y_{t_2})}{\log\rho}$ and it is sufficient to consider the interference unknown part of the received signal at both destinations as an $N$-to-$1$ MISO channel with capacity $I(x_{s_1},x_{s_2};y_{t_1})=I(x_{s_2};y_{t_1})+I(x_{s_1};y_{t_1}|x_{s_2})$ to give a DMT lower bound. In the high SNR regime, the exponential order of $\beta_{r_n}$ vanishes and thus we have
\begin{eqnarray}
&\quad&\lim_{\rho\rightarrow\infty}\frac{I(x_{s_2};y_{t_1})}{\log\rho} \notag \\
&=&\lim_{\rho\rightarrow\infty}\frac{\log(1+\frac{\rho\sum_{n=1}^N|h_{s_2,r_n}|^2|\beta_{r_n}|^2|h_{r_n,t_1}|^2}{\rho\sum_{n=1}^N|h_{s_1,r_n}|^2|\beta_{r_n}|^2|h_{r_n,t_1}|^2+1})}{\log\rho}\notag\\
&=&[\max_{n=1,2,...,N}\{1-v_{2,n}-u_{n,1}\}-(\max_{n=1,2,...,N}\{1-v_{1,n}-u_{n,1}\})^+]^+
\label{eqn:dmtsecond1}.
\end{eqnarray}
From the definition of the outage probability, we know that
\begin{eqnarray}
P_{O}&\doteq&P[I(x_{s_1};y_{t_1})+I(x_{s_2};y_{t_2}|x_{s_1})<R-R_{t_1}+R-R_{t_2}] \notag\\
&=&P[\log(1+\frac{\rho\sum_{n=1}^N|h_{s_1,r_n}|^2|\beta_{r_n}|^2|h_{r_n,t_1}|^2}{\rho\sum_{n=1}^N|h_{s_2,r_n}|^2|\beta_{r_n}|^2|h_{r_n,t_1}|^2+1})\notag\\ &\quad&+\log(1+\rho\sum_{n=1}^N|h_{s_2,r_n}|^2|\beta_{r_n}|^2|h_{r_n,t_2}|^2)<(r^c_{t_1}+r^c_{t_2})\log\rho],
\label{eqn:outagesecond}
\end{eqnarray}
where $r^c_{t_1}$ denotes the multiplexing gain of $x_{s_1}$ at destination node $t_1$ with $R-R_{t_1}$ amount of information and unknown interference signal and $r^c_{t_2}$ is similarly defined. From \eqref{eqn:outageorder2}, \eqref{eqn:dmtsecond1} and \eqref{eqn:outagesecond}, the outage events set $O^+$ should be defined as
\begin{eqnarray}
O^+&=&\{(\textbf{v},\textbf{u})\in\mathbb{R}^{3N+}|[\max_{n=1,2,...,N}\{1-v_{2,n}-u_{n,1}\} \notag \\
&\quad&-(\max_{n=1,2,...,N}\{1-v_{1,n}-u_{n,1}\})^+]^++[\max_{n=1,2,...,N}\{1-v_{1,n}-u_{n,1}\}]^+<r^c_{t_1}+r^c_{t_2}\}.
\label{eqn:outageeventssecond}
\end{eqnarray}
Thus, in order for the outage events to happen, the following constraints must be satisfied:
\begin{eqnarray}
v_{2,n}+u_{n,1}>1-r^c_{t_1}-r^c_{t_2}, \forall{} n=1,2,...,N.
\end{eqnarray}
From \eqref{eqn:do} and \eqref{eqn:outageeventssecond}, we can lower-bound the DMT of the interference unknown part of the received signal at both destinations $d_{II}(r^c_{t_1},r^c_{t_2})$ as
\begin{eqnarray}
d_{II}(r^c_{t_1},r^c_{t_2})&\geqslant&d_{O}(r^c_{t_1},r^c_{t_2})\notag\\
&\geqslant&\inf_{(\textbf{v},\textbf{u})\in{O^+}}[\sum_{n=1}^N(v_{1,n}+u_{n,1}+v_{2,n})] \notag \\
&>&N(1-r^c_{t_1}-r^c_{t_2}).
\end{eqnarray}
Because $r^c_{t_1}=\frac{R-R_{t_1}}{R}r$ and $r^c_{t_2}=\frac{R-R_{t_2}}{R}r$, using the same technique from \eqref{eq:dmtfirst} to \eqref{eq:dmt24}, we can further write
\begin{eqnarray}
d_{II}(r^c_{t_1},r^c_{t_2})>N[1-(\frac{R}{R-R_{t_1}}+\frac{R}{R-R_{t_2}})r].
\label{eq:dmt2}
\end{eqnarray}

\subsubsection{Overall result}
From \eqref{eqn:dmt1}, \eqref{eq:dmt2} and \eqref{eqn:overalldmt}, we know the overall achievable DMT for the active PNC strategy is lower-bounded as
\begin{eqnarray}
d(r)>N[1-\frac{2R(2R-R_{t_1}-R_{t_2})}{2R^2-R^2_{t_1}-R^2_{t_2}}r].
\end{eqnarray}
\begin{figure}
    \centering
    \includegraphics[width=14cm]{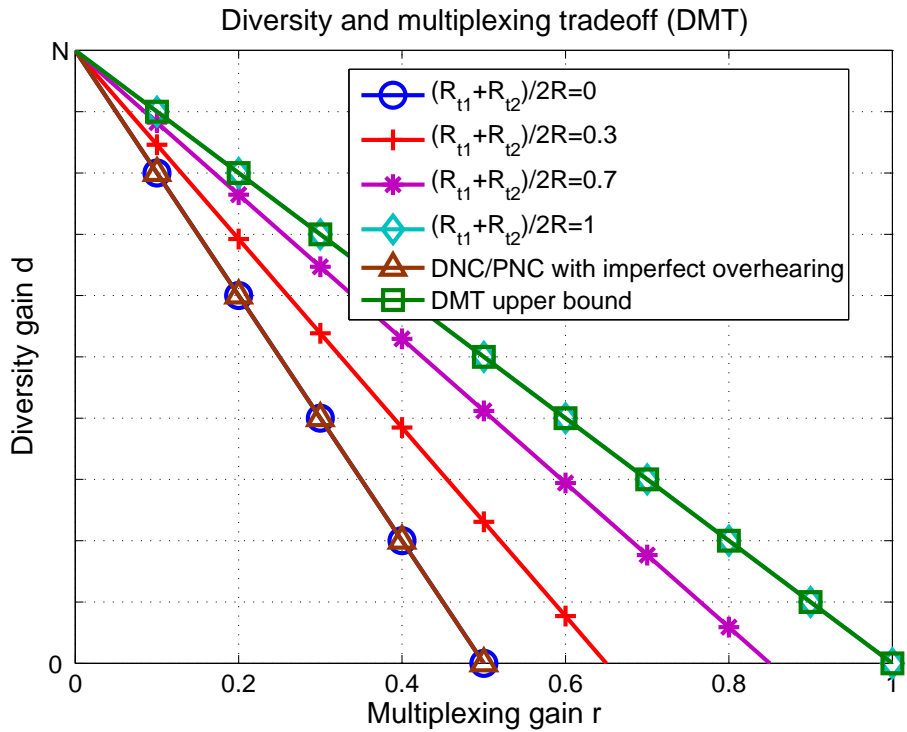}
    \caption{DMT of active PNC strategy with imperfect overhearing.}
    \label{fig:dmt}
\end{figure}

As shown in Fig. \ref{fig:dmt}, the more information overheard by the destinations, the better DMT the active PNC strategy can achieve. In the case of perfect overhearing, it can achieve the DMT upper bound, which is obtained by viewing the channel model as two two-hop fully cooperative MISO channels; in the case of imperfect overhearing, DNC and PNC can only achieve the DMT of traditional multihop routing, while the active PNC strategy can still improve the overall system performance by using the correctly overheard information. This indicates that, although introducing bi-directional interference to both destinations due to the imperfect overhearing, the active PNC strategy's actively mixing the two source nodes' signal at the relay nodes, regardless the reception quality of the overheard information, in general improves the overall system throughput and robustness.
\section{Conclusion}
In this paper, we studied the problem that whether or not we should still use network coding to improve the system performance if the overheard information is imperfect. DNC and PNC strategies fall back to traditional multihop routing strategy when the overhearing is imperfect. While it is clearly huge waste to ignore the whole overheard information with only a few incorrect symbols in it, we proved that even when the overhearing is imperfect, an active network coding strategy could still push the DMT closer to the upper bound. This tells us that in general wireless networks, where overhearing is rich but lossy, an adaptive strategy that switches between network coding strategy and traditional multihop routing strategy is often not optimal. Thus, a better communication strategy is to use network coding as much as possible regardless of the reception quality of the wireless overhearing.
\bibliographystyle{IEEEtran}
\bibliography{haishining}
\end{document}